\documentclass[%
 reprint,
 superscriptaddress,
 amsmath,amssymb,
 aps,
 prl,
]{revtex4-1}

\usepackage{graphicx}
\usepackage{dcolumn}
\usepackage{bm}
\usepackage{xcolor}
\usepackage{braket}
\DeclareMathOperator{\tr}{tr}
\DeclareMathOperator{\erf}{erf}

\usepackage[textsize=tiny]{todonotes}

\begin{document}

\preprint{APS/123-QED}

\title{Avoided quantum criticality in exact numerical simulations of a single disordered Weyl cone}

\author{Justin H. Wilson}
\affiliation{Department of Physics and Astronomy, Center for Materials Theory, Rutgers University, Piscataway, NJ 08854 USA}
\author{David A. Huse}
\affiliation{Physics Department, Princeton University, Princeton, New Jersey 08544, USA}
\affiliation{Institute for Advanced Study, Princeton, New Jersey 08540, USA}
\author{S. Das Sarma}
\affiliation{Condensed Matter Theory Center and Joint Quantum Institute, Department of Physics, University of Maryland, College Park, MD 20742, USA}
\author{J. H. Pixley}
\affiliation{Department of Physics and Astronomy, Center for Materials Theory, Rutgers University, Piscataway, NJ 08854 USA}

\date{\today}

\begin{abstract}
Existing theoretical works differ on whether three-dimensional Dirac and Weyl semimetals are stable to a short-range-correlated random potential.
Numerical evidence suggests the semimetal to be unstable, while some field-theoretic instanton calculations have found it to be stable. 
The differences go beyond method: the continuum field-theoretic works use a single, perfectly linear Weyl cone, while numerical works use tight-binding lattice models which inherently have band curvature and multiple Weyl cones.
In this work, we bridge this gap by performing exact numerics on the same model used in analytic treatments, and we find that all phenomena associated with rare regions near the Weyl node energy found in lattice models persist in the continuum theory: The density of states is non-zero and exhibits an avoided transition.
In addition to characterizing this transition, we find rare states and show that they have the expected behavior.
The simulations utilize sparse matrix techniques with formally dense matrices; doing so allows us to reach Hilbert space sizes upwards of $10^7$ states, substantially larger than anything achieved before.
\end{abstract}

\maketitle

The stability of phase transitions in the presence of non-perturbative effects of rare regions is a central question in modern statistical mechanics~\cite{Vojta-2006,Agarwal-2017,Syzranov-2018}. 
These problems fall into two classes; the first is the case of ``clean'' critical point perturbed by disorder, and the second consists of transitions driven solely by disorder. 
The latter case is less understood as both the existence of the transition and the rare region effects arise from the same origin: randomness. 
As a result, rare regions could destabilize one of the two phases turning a putative transition into a crossover.

The problem of three-dimensional short-range disordered Dirac and Weyl semimetals~\cite{Armitage-2017} is a quintessential example of a disorder-driven transition that has a non-trivial interplay with non-perturbative, rare-region effects
~\cite{Fradkin-1986,Goswami-2011,Kobayashi-2014,Brouwer-2014,Bitan-2014,*Bitan-2016,Nandkishore-2014,Pixley-2015,Altland-2015,Sergey-2015,Leo-2015,Sbierski-2015,Pixley2015disorder,Garttner-2015,Liu-2015,Bera-2015,Shapourian-2015,Altland2-2015,Sergey2-2015,Louvet-2016,Pixley-2016,Pixley2,Sbierski-2017,Pixley-2017,Guararie-2017,Holder-2017,Wilson-2017,Wilson-2018,Ziegler-2018,Buchhold-2018,*Buchhold2-2018,Balag-2018,Syzranov-2018,Klier-2019}. 
Initial work using large-$N$~\cite{Fradkin-1986} and perturbative renormalization group~\cite{Goswami-2011} found that Dirac and Weyl semimetals are stable to the presence of weak disorder and possess a quantum phase transition into a diffusive metal phase; this is indicated by the order parameter, the density of states at zero energy, becoming non-analytic at the transition.  
On the other hand, rare-region arguments and mean-field instanton calculations~\cite{Nandkishore-2014} argued that non-perturbative effects lead to a finite density of states at the Weyl (or Dirac) node for infinitesimal disorder strength, thus destabilizing the semimetallic phase.

Confirming the rare region expectation, extensive numerical simulations on lattice models of Dirac and Weyl semimetals have found non-perturbative rare eigenstates that round the perturbative transition into a crossover dubbed an avoided quantum critical point (AQCP)~\cite{Pixley-2016,Pixley2,Pixley-2017,Wilson-2017,Wilson-2018} with an analytic density of states. 
A phenomenological, field-theoretic description of the AQCP has been put forth~\cite{Guararie-2017}, and additional support for the rare region scenario comes from $T$-matrix calculations of the quasiparticle lifetime~\cite{Pixley-2017}, conductivity~\cite{Holder-2017}, and the prediction of a non-zero density of states from a continuous distribution of scattering approach~\cite{Ziegler-2018}. 
Last, by replacing the randomness by quasiperiodicity, rare regions are removed entirely from the problem, and a genuine quantum phase transition between a Weyl semimetal and a diffusive metal is seen~\cite{Pixley-2018}, albeit with no randomness in the model.

\begin{figure}
    \centering
    \includegraphics[]{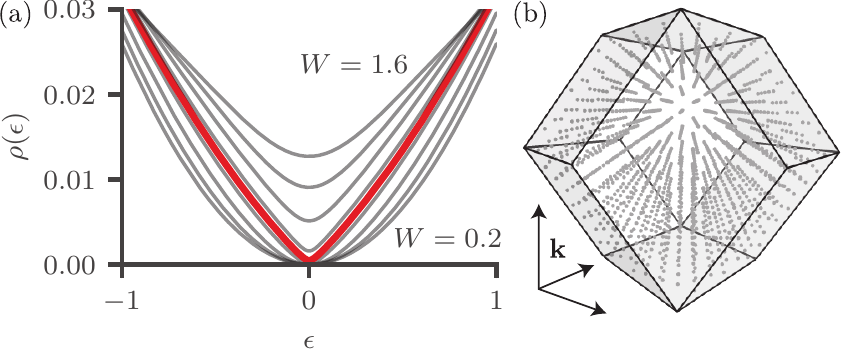}
    \caption{(a) Density of states $\rho(\epsilon)$ as a function of energy $\epsilon$ for various different values of disorder $W$ ranging from $W = 0.2$ to $ W = 1.6$ in steps of $0.2$. The red curve is for $ W = 0.9$ and is close to the avoided transition.  Far enough away from zero energy this behaves as $\rho(\epsilon)\sim |\epsilon|$, consistent with avoided criticality. (b) Depiction of the FCC momentum space lattice and the rhombic dodecahedron Brillouin zone.}
    \label{fig:1}
\end{figure}

Recently, continuum field-theoretic work that considered fluctuations about the instanton saddle point for a single Weyl cone have challenged the rare region scenario~\cite{Buchhold-2018,*Buchhold2-2018}. 
In Ref.~\onlinecite{Buchhold-2018,*Buchhold2-2018} the authors find that while rare regions exist, they do not destabilize the semimetallic phase because the density of states at the Weyl node remains zero for non-zero disorder. An immediate conclusion of this scenario is that the perturbative quantum critical point remains stable to disorder; in the present work, we directly investigate this question in a numerical realization of a single Weyl cone. 
Previous simulations~\cite{Pixley-2016,Pixley2,Pixley-2017,Wilson-2017,Wilson-2018} have at least two Weyl cones (due to the fermion doubling theorem~\cite{Nielsen81PLB,*Nielsen81NPB1,*Nielsen81NPB2}), internode scattering, and band curvature effects; their conclusions, strictly speaking, do not apply to a single Weyl cone with a linear dispersion at all energies. While some numerical results exist on the conductance in the limit of a single Weyl cone~\cite{Brouwer-2014,Sbierski-2015}, no rare region effects have been reported. 
Additionally, the existing numerical techniques that have been highly successful in reaching large enough system sizes to observe rare region effects rely on sparse matrices (that naturally occur in local lattice models) and efficient matrix-vector multiplication not directly applicable to treat disordered continuum models.
Therefore, the important issue remains open on whether the conflicting conclusions about the existence of QCP versus AQCP have perhaps been obtained in different models.
To resolve this, we employ exact numerics which necessarily include disorder realizations past those considered in Ref.~\cite{Buchhold-2018,*Buchhold2-2018} 
as they can contain numerous rare-regions~\cite{Pixley-2016}.

In the present Rapid Communication, we numerically study a disordered, single Weyl node by adapting sparse matrix-vector routines to work in the continuum. 
Technically, we achieve this by using fast Fourier transforms to act with the disorder potential in its diagonal real space basis (similar to Ref.~\onlinecite{sbierskiStrongDisorderNodal2019}). 
Importantly, the relevant sparse matrix algorithms which scale with (single-particle) Hilbert space dimension $\mathcal{N}$ only increases from $\mathcal{N}$ to $\mathcal{N}(\log \mathcal{N})^3$ in three dimensions. 
To treat the continuum limit, we consider two controlled ways to discretize momentum space. 
First, we demonstrate the existence of rare regions in a model of a disordered, single Weyl cone. 
Second, we study the density of states near the Weyl node. We demonstrate avoidance of the perturbative transition; the density of states near this avoided transition is finite and remains an analytic function of energy and disorder near the Weyl node.

In Fig.~\ref{fig:1}(a), we show an example of the density of states $\rho(\epsilon)$ as a function of energy $\epsilon$ and disorder strengths $W$ across the AQCP $W_c$. 
Approaching the AQCP, the $\rho(\epsilon)$ scaling goes from $\sim \epsilon^2$ to $\sim |\epsilon|$ scaling at the (avoided) transition, consistent with the renormalization group expectation ($z=3/2$). 
However, this scaling does not persist to zero energy due to the non-zero density of states at $\epsilon =0$. 
As we track the zero energy density of states $\rho(0)$ for $W<W_c$, we find that $\rho(0)$ is non-zero (converged with system size) and decreases in an exponential fashion, thus ruling out the stability of the semimetal phase.  
Importantly, all of our conclusions are unaffected by the discretization of the continuum.
We conclude that AQCP survives the continuum single-cone limit.

\emph{Continuum model and numerical implementation}:
The model for a single disordered Weyl cone takes the form
\begin{equation}
    \mathcal H = -i \hbar v_{F} \bm \sigma \cdot \bm{\nabla} + V(\mathbf r), \; \braket{V(\mathbf r + \mathbf R) V(\mathbf{r}) } = W^2 e^{-\frac{R^2}{\xi^2}},
\end{equation}
where $\braket{\cdots}$ represents the disorder average, and $V(\mathbf r)$ is a Gaussian random variable with zero mean.
Without loss of generality, we take $v_{F} = 1 = \hbar$ and $\xi=1$.
For simulation purposes, we use the momentum space version of the problem where
\begin{equation}
    \mathcal{H}_{\mathbf k,\mathbf k'} = \bm \sigma \cdot \mathbf k (2\pi)^3 \delta(\mathbf k - \mathbf k') + V(\mathbf k - \mathbf k'),
\end{equation}
 the Gaussian disorder in the potential takes the form \cite{supp}
\begin{equation}
    \braket{V(\mathbf k) V(\mathbf k')^*} = W^2 \pi^{3/2} e^{-\frac14 k^2} (2\pi)^3 \delta(\mathbf k - \mathbf k'), \label{eq:momentum-space-correlator}
\end{equation}
and we define the Fourier transform such that $\Psi(\mathbf k) = \int d^3 x \, \Psi(\mathbf x) e^{-i\mathbf k \cdot \mathbf x}$.

To discretize the problem, we construct a grid in momentum space characterized by three lattice vectors $\mathbf b_j$ defined as columns of a matrix $B$. 
Momentum is found by a vector of integers $\mathbf n$ via $\mathbf k_{\mathbf n} = \delta k B( \mathbf n + \bm \varphi)$ where $\delta k = \frac{2\pi}{N a}$ for length scale $a$, number of grid points with linear dimension $N$, and offset $\bm \varphi \in [0,1)^3$.
The length scale $a$ is related to a real space lattice spacing,  $Na$ to a system-size, and we have periodic boundary conditions in real space. 
We consider two different momentum space lattices: cubic and face-centered cubic (FCC). The FCC lattice provides the densest packing of spheres in three-dimensions, allowing us to approximate the continuum more accurately for a given number of momentum-space grid points.
For a cubic (momentum-space) lattice $B$ is the identity, $a$ is the lattice constant, and $L = aN$ is the system size, but for a FCC lattice, $B_{ij} = 1/2$ if $i\neq j$ and $B_{ii} = 0$, the real-space lattice is body-centered cubic (BCC) occupying a rhombohedron with side-length $L = \sqrt{3} a N$ and angle between sides $\alpha = \arccos(-1/3)$.
Similarly, the constructed grid determines the momentum space cut-off $\Lambda$ by half the size of the linear dimension. 
For cubic discretization, $\Lambda = \frac{\pi}a$ with a cubic cutoff around $\mathbf k=0$ while for the FCC discretization $\Lambda = \frac{\pi}{\sqrt{2} a}$ with a rhombic dodecahedron cutoff around $\mathbf k=0$, as depicted in Fig~\ref{fig:1}(b). 

To discretize the Hamiltonian, we consider its action on a wave function
\begin{equation}
    \mathcal{H}\Psi = \int \frac{d^3 k'}{(2\pi)^3}\mathcal{H}_{\mathbf k, \mathbf k'} \Psi(\mathbf k') \approx \sum_{\mathbf{n}'} H_{\mathbf n\mathbf n'} \psi_{\mathbf n'},
\end{equation}
where $H_{\mathbf n\mathbf n'} \equiv \frac{\det(B)}{(Na)^3} \mathcal{H}_{\mathbf{k}_{\mathbf{n}},\mathbf{k}_\mathbf{n}'}$  and $\psi_{\mathbf n} \equiv \Psi(\mathbf k_{\mathbf n})$.
Discretization affects the Dirac delta function such that
$\delta(\mathbf k_{\mathbf n} - \mathbf k_{\mathbf{n}'}) \approx \tfrac{(N a)^3}{(2\pi)^3\det(B)} \delta_{\mathbf n \mathbf n'}$. 
This simplifies the kinetic term and discretizes the correlator in Eq.~\eqref{eq:momentum-space-correlator}
\begin{equation}
    \braket{V_{\mathbf n} V_{\mathbf n'}^*} = W^2 \pi^{3/2} e^{-\frac{1}4(\delta k B\mathbf{n})^2} \tfrac{(N a)^3}{\det(B)} \delta_{\mathbf n \mathbf n'},
\end{equation}
where $V_{\mathbf n} \equiv V(\delta k B \mathbf n)$.
This correlator is achieved by \cite{Chou-2014}
\begin{equation}
    V_{\mathbf{n}} = W z_{\mathbf n} \sqrt{\tfrac{(N a)^3}{\det(B)} \pi^{3/2}  e^{-\frac{1}4(\delta k B\mathbf{n})^2} },
\end{equation}
where $z_\mathbf{n}$ are Gaussian i.i.d.\ random complex numbers with $\braket{z_{\mathbf{n}}} = 0$ and $\braket{z_{\mathbf{n'}}^*z_{\mathbf{n}}} = \delta_{\mathbf{n}\mathbf{n'}}$.
To ensure $V$ is hermitian, we find the inversion operator for our lattice $P$ and identify $z_{\mathbf{n}} = z^*_{P\mathbf{n}}$ and make sure inversion symmetric points are real-valued. 
We also impose $V_{\mathbf{0}} = 0$ to avoid random spatially-uniform shifts in the potential.

The discretized Hamiltonian is
\begin{equation}
    H_{\mathbf n \mathbf n'} = \delta k\,  \bm \sigma \cdot B(\mathbf n + \bm \varphi)\delta_{\mathbf n \mathbf n'} + \frac{\det(B)}{(N a)^3}  V_{\mathbf n - \mathbf n'}.
\end{equation}
This matrix as written is \emph{dense}, but to take advantage of numerical techniques that only require matrix-vector multiplication, we  consider how this acts on a vector $\psi_{\mathbf n}$.
First, the kinetic part is block diagonal, but the potential acts as a convolution.
To implement a convolution, we need the three-dimensional Fourier transform of $V_{\mathbf n}$.
The result is a linear operator
\begin{equation}
    \frac{\det(B)}{(a N)^3}\sum_{\mathbf n'} V_{\mathbf n- \mathbf n'}\psi_{\mathbf n'} = \frac{\det(B)}{a^3} \mathcal F\bm{[}\mathcal F^{-1}[V_{\mathbf{n}}] \mathcal F^{-1}[\psi_{\mathbf{n}}]\bm{]},
\end{equation}
where $\mathcal F$ is a three-dimensional fast Fourier transform (FFT).
The FFT is, in a sense, returning us to real space where the potential is diagonal, but for our purposes, we consider it a tool for the application of the convolution. As Lanczos and the kernel polynomial method (KPM)~\cite{Weisse-2006} based approaches for sparse matrices scale like $\sim \mathcal{N}$ (for matrix size $\mathcal{N}$) the inclusion of the FFT only increases the computational cost to $\mathcal{N}(\log \mathcal{N})^3$, which keeps the algorithm sufficiently fast. Thus, our approach provides an efficient way to utilize matrix-vector routines to study inhomogeneous continuum models.

Using an FFT introduces a notion of Brillioun zones (BZs).
For any finite BZ there is a discontinuity in the kinetic energy at the edge of the BZ due to the Fermion doubling theorem: There ought to be a second Weyl fermion at the BZ edge with infinite velocity (but our finite grid never picks it up).
Further, the convolution acts across the BZ, connecting $\mathbf k$ points that are far from each other in the continuum but close in a periodic BZ.
We expect that this only affects the high energy behavior and does not affect the low energy regime of interest that we are probing near $E=0$.
To confirm this, we compare two models with different cutoff physics (1) the cubic lattice and (2) the FCC lattice, and we find that there is qualitatively no difference in the low energy physics we study \cite{supp}.
We illustrate the FCC lattice in Fig.~\ref{fig:1}(b). 

\begin{figure}[t!]
    \centering
    \includegraphics[]{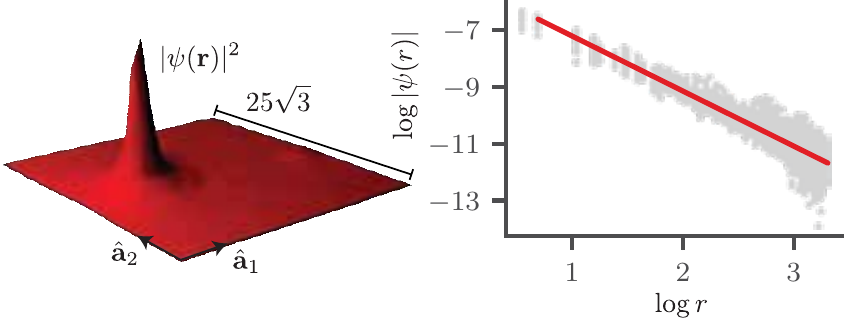}
    \caption{{\bf Properties of a rare wavefunction computed with the FCC model with $L = 25\sqrt{3}$, $\Lambda = \tfrac{\pi}{\sqrt{2}}$}, {\bf and a disorder strength below the AQCP} $W=0.7<W_c(\Lambda)\approx 0.9$. The energy of the state is $\epsilon = 0.0168$ but can made to pass smoothly through zero energy with a small perturbation of the disorder potential~\cite{supp}. (left) The probability density of a rare wavefunction for a cut through the real space BCC lattice 
    where $\hat{\mathbf a}_1$ [$\hat{\mathbf a}_2$] is in the $(1,-1,1)$ [$(-1,1,1)$] direction and  $\mathbf r = n_1 \mathbf{a}_1 + n_2 \mathbf{a}_2 + 22 \mathbf{a}_3$. (right) A scatter plot of the wavefunction as a function of the distance to its maximum value demonstrating a clear power law decay with $|\psi(\mathbf r)| \sim 1/r^{1.94}$ for this rare state. }
    \label{fig:2}
\end{figure}

Finally, if we stochastically sample $\bm\varphi$, we reproduce the continuous density of states for the continuum system; all finite size effects are then from the discretization of $V(\mathbf k)$.
Physically, a nonzero $\bm \varphi$ is usually associated with twisted boundary conditions in real space.

\begin{figure*}[t!]
    \centering
    \includegraphics{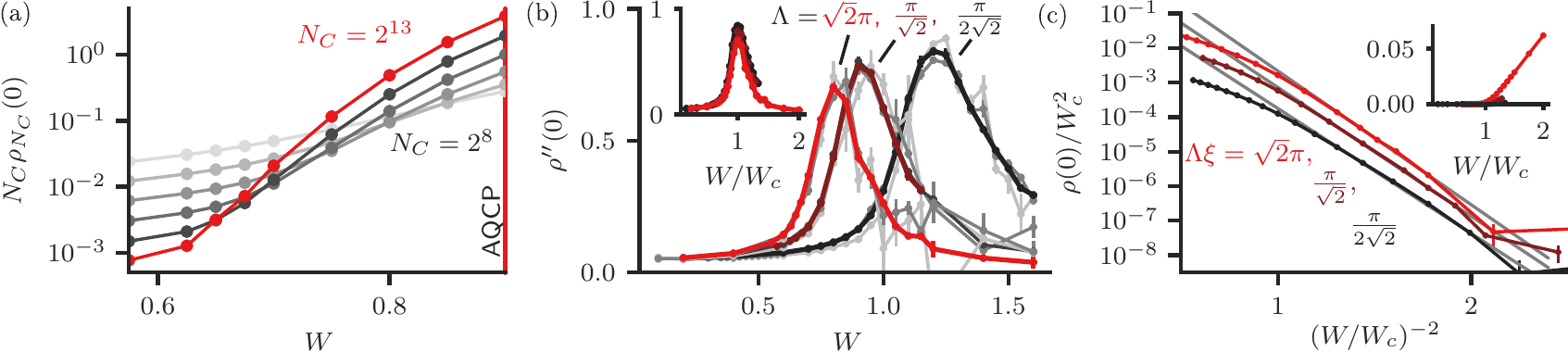}
    \caption{{\bf Demonstrating the avoided transition and finite density of states.} (a) From lightest to darkest each curve represents $N_C=2^8$ to $2^{13}$ in multiples of two ($\Lambda=\pi/\sqrt{2}$ and $L=160\sqrt{3}$).
    Other measures of the avoided criticality put $W_c = 0.9$ but we see that there is no saturation of $N_C\rho_{N_C}(0)$ as we would expect from scaling with $N_C$ at the quantum critical point. 
    (b) We fit $\rho_{N_C}(0)$ to Eq.~\eqref{eq:NCscaling} to extract $\rho(0)$ and $\rho''(0)$ in the $N_C\rightarrow\infty$ limit~\cite{supp}. We find for each cutoff studied, a peak in $\rho''(0)$  saturated in system size (all gray curves are smaller system sizes $L \Lambda\sqrt{2}/\pi =32\sqrt{3},64\sqrt{3},128\sqrt{3}$ and for $\Lambda=\pi/\sqrt{2}$ we include $L = 160\sqrt{3}$). (b,inset) Normalized with respect to the peak value ($W_c = 0.8,0.9,1.2$ left to right), the peaks line up well with each other, displaying a weak cut-off dependence. (c) We find  $\rho(0)$ is well fit by the rare region form in Eq.~\eqref{eqn:dosRR} over $\approx 4$ orders of magnitude (grey lines). The data as plotted are converged in system size, as the gray curves clearly indicate in (b) and as expanded upon in the Supplement~\cite{supp} for (c). The inset shows the same data on a linear scale.
    }
    \label{fig:3}
\end{figure*}

Defining $H$ as a linear operator allows us to take advantage of numerical techniques that only involve matrix-vector multiplication such as Lanczos and the KPM~\cite{Weisse-2006}.
Lanczos is used to obtain eigenvectors near zero energy $\epsilon=0$, and we obtain averaged density of states
\begin{equation}
    \rho_{\mathrm{dis}}(\epsilon) = \bigg\langle{\frac1{2N^3} \sum_{n} \delta(\epsilon - \epsilon_n)}\bigg\rangle,
\end{equation}
with the KPM.
To relate the density of states of the discretized Hamiltonian to its continuum counterpart, a measure factor is required from $d^3 k \approx (\delta k)^3\det(B)$ which leads to  $\rho(\epsilon) = \frac{\det(B)}{a^3} \lim_{L,\Lambda \rightarrow\infty}  \rho_{\mathrm{dis}}(\epsilon)$ for fixed $W$.

The KPM method uses a Chebyshev expansion to order $N_C$, leading to a density of states $\rho_{N_C}(\epsilon)$ \cite{supp} which behaves as a convolution of the exact $\rho_{\mathrm{dis}}(\epsilon)$ with a Gaussian of width $\delta \epsilon = \frac{\pi\Delta}{N_C}$ and bandwidth $\Delta$ of $H$.
We probe the scaling of $\rho_{N_C}(0)$ with $N_C$ to asses the low energy behavior of $\rho(\epsilon)$.
Precisely, assuming the density of states is analytic, we Taylor expand $\rho(\epsilon)$ to find
\begin{equation}
   \rho_{N_C}(0) = \rho(0) + \tfrac12 \rho''(0) \left(\tfrac{\pi\Delta}{N_C} \right)^2 + \cdots , \label{eq:NCscaling}
\end{equation}
and at the perturbative critical point, if we have $\rho(\epsilon) \sim |\epsilon|$, then  
\begin{equation}
    \rho_{N_C}(0) \sim \tfrac1{N_C}\text{ and } \rho''_{N_C}(0) \sim N_C.  \label{eq:criticalNCscaling}
\end{equation}
We also numerically compute $\rho_{N_C}''(0)$ directly from the KPM expansion~\cite{Pixley2}.

\emph{Finding rare states}: We begin by finding a low-energy rare state in the weak disorder regime, i.e., below the avoided transition. 
We use Lanczos on $H^2$ to find states that are not in the perturbative ``Dirac peaks''~\cite{Pixley-2016}; such an example is shown in Fig.~\ref{fig:2} that is power-law bound to the region (at ${\bf r}_0$) of uncharacteristically high disorder strength. 
The rare wavefunction decays like $\psi({\bf r}) \sim 1/|{\bf r}-{\bf r}_0|^{\alpha}$, where $\alpha = 1.94$ in excellent agreement with the analytic prediction at the saddle point $\alpha=2$. 
In summary, the rare wavefunction we have found here shares all of the same characteristics as in lattice model simulations, and we find that they are not any more difficult to find.

\emph{Behavior of the density of states}:
We now turn to a detailed analysis of the density of states. 
To get accurate results we average over a large number of disorder samples ranging from 2,500 to 25,000 and analyze the zero energy density of states following Eq.~\eqref{eq:NCscaling} to extract $N_C$-independent estimates of $\rho(0)$ and $\rho''(0)$~\cite{supp}. 
We use $\rho''(0)$ to determine whether the density of states becomes non-analytic, which would imply $\rho''(0) \rightarrow \infty$. 
In addition to the $N_C$-independent estimate of $\rho''(0)$ we also compute it directly at fixed $N_C$ within the KPM~\cite{Pixley2} which we denote as $\rho_{N_C}''(0)$.
Note that, if the critical point exists it implies the scaling $\rho_{N_C}(0)\sim 1/N_C$ and $\rho''_{N_C}(0)\sim N_C$. 
We test for this scaling by plotting $N_C \rho_{N_C}(0)$ vs.\ $W$, if it holds then different $N_C$ curves should intersect at \emph{one common point}. 
However, as shown in Fig.~\ref{fig:3}(a), we find that no such crossing occurs; instead, each increasing pair of $N_C$'s intersects at smaller values of $W$, indicating the absence of a transition at the lowest energy scales. 

Our second piece of evidence for the avoided transition is the strongly rounded peak in $\rho''(0)$, as shown in Fig.~\ref{fig:3}(b). We find that $\rho''(0)$ is converged in system size (gray curves indicate smaller system sizes), not singular, and weakly dependent on the cut-off. 
Thus, we find that the density of states remains an analytic function of $W$ and $\epsilon$ at the Weyl node, except for an expected essential singularity at $W=0$ due to the nonperturbative disorder effects. 
The location of the maximum of the peak provides an accurate estimate of the avoided transition $W_c(\Lambda)$~\cite{Pixley-2016,Pixley2} that also agrees with the estimate based on the apparent scaling $\rho(\epsilon)\sim |\epsilon|$.

\begin{figure}[b!]
    \centering
    \includegraphics[]{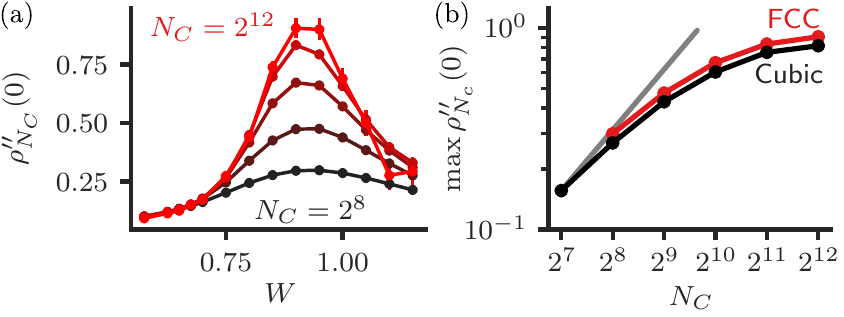}
    \caption{{\bf Convergence of the peak of $\rho_{N_C}''(0)$ at the AQCP}. (a) $N_C$-dependence of the second derivative of the density of states for the FCC model. (b) A plot of the peak saturated $\rho''_{N_C}(0)$ data for both FCC and cubic data. Data at fixed $N_C$ are converged in system size \cite{supp}.  We see that they do not match the scaling we would expect from a true quantum critical point $\rho''_{N_C}(0) \propto N_C$ (grey line). The FCC data has $\Lambda = \pi/\sqrt{2}$ and $L = 160\sqrt{3}$ while the cubic data has $\Lambda = \pi$ and $L=160$. }
    \label{fig:4}
\end{figure}

Tracking the zero energy density of states for decreasing $W$ below the avoided transition, we converge the $N_C$-independent $\rho(0)$  in system size~\cite{supp} to an exponentially small but non-zero value. As shown in Fig.~\ref{fig:3}(c), we find that the converged value of $\rho(0)$ is well described by the results of the saddle point instanton expectation~\cite{Nandkishore-2014},
\begin{equation}
     \rho(0) = a(\Lambda)\exp\left[ -b\left(W_c(\Lambda)/W\right)^2\right].
    \label{eqn:dosRR}
\end{equation}
Impressively, the data fit to this form extends over three to four orders of magnitude in $\rho(0)$ depending on the cut-off. 
We find that all of the results share a common slope (with the fitted value ranging from $b=8.3 \pm 0.7$ to $b = 8.9 \pm 1.1$ where the error is mostly due to $W_c$ error) and the offset (i.e. the prefactor $a(\Lambda)$) is cut-off dependent.  
Thus, as $\rho(0)$ is converged in system size [Fig.~\ref{fig:3}(c)] and $N_C$, it is finite in the thermodynamic limit and increases with increasing cut-off. 
These results imply that rare regions have induced a non-zero density of states for any finite value of $W$ and below the avoided transition,  Eq.~\eqref{eqn:dosRR} describes it.

Finally, we present  $\rho_{N_C}''(0)$ directly computed from the KPM expansion to demonstrate any rounding from performing fits to Eq.~\eqref{eq:NCscaling} is weak and the lack of divergence of $\rho''(0)$ is \emph{intrinsic} to the problem. 
As shown in Fig.~\ref{fig:4}(a) we find that the peak in $\rho_{N_C}''(0)$ grows with $N_C$ but at the largest $N_C$'s the increase is minor, demonstrating saturation with $N_C$. 
For clarity and to test the critical scenario $\rho_{N_C}''(0)\sim N_C$, we plot the peak value of $\rho_{N_C}''(0)$ as a function of $N_C$ in Fig.~\ref{fig:4}(b). 
We find that the peak is saturating with $N_C$ (independent of the kind of discretization of the continuum) and does not come close to the critical scaling expectation. 
It is useful to contrast the rise in $\rho_{N_C}''(0)$ with the quasiperiodic limit of the model~\cite{Pixley-2018}, which has an actual transition and the divergence in $\rho_{N_C}''(0)$ manifests as an increase over six orders of magnitude.  In contrast, in the present model, the peak barely rises over one order of magnitude. The transition is strongly avoided.

\emph{Discussion:} The physical role of rare states in causing a nonzero density of states appears unchanged from lattice models~\cite{Pixley-2016,Pixley2,Pixley-2017,Wilson-2017,Wilson-2018}, and stands in contrast to analytic results~\cite{Buchhold-2018,Buchhold2-2018}.
The analytic work suggests that a single spherical potential (e.g.\ one rare state) cannot lead to a density of states at zero energy.
However, the exact numerics have configurations with multiple rare states, which produce long-range tunneling matrix elements that fall of like their distance between them squared~\cite{Nandkishore-2014,Pixley-2016}.
On the other hand, to make a direct comparison of a single rare event [Fig.~\ref{fig:2}] and those of Ref.~\cite{Buchhold-2018,Buchhold2-2018}
the vector of angles $\bm \varphi$ (used in simulations) ought to be physically considered; it can be mapped \emph{exactly} to a Bloch wave vector where the disorder potential on an $N\times N \times N$ lattice is repeated infinitely in space.
In this paradigm, even single rare states map to an infinite band of rare states, and the saturation of density of states with system size indicates that this band does not become sparser for larger supercells (i.e.\ larger $N$). 
Therefore, the density of rare states participating in these bands remains constant.

\acknowledgements{
\emph{Acknowledgements}: 
We thank Alexander Altland and Michael Bucchold for illuminating conversations.
D.A.H.\ was supported in part by a Simons Fellowship and by DOE grant DE-SC0016244.
S.D.S.\ is supported by the Laboratory for Physical Sciences and Microsoft.
J.H.P.\ is supported by NSF CAREER Grant
No. DMR-1941569.
The authors acknowledge the following research computing resources that have contributed to the results reported here: 
The University of Maryland supercomputing resources (http://hpcc.umd.edu),
the Beowulf cluster at the Department of Physics and Astronomy of Rutgers University; and the Office of Advanced Research Computing (OARC) at Rutgers, The State University of New Jersey (http://oarc.rutgers.edu), for providing access to the Amarel cluster.  }

\bibliography{DSM_RR}

\clearpage
\pagebreak
\widetext
\begin{center}
    \textbf{\large Supplement to ``Avoided quantum criticality in exact numerical simulations of a single disordered Weyl cone''}
\end{center}

\setcounter{equation}{0}
\setcounter{figure}{0}
\setcounter{table}{0}
\setcounter{page}{1}
\renewcommand{\theequation}{S\arabic{equation}}
\setcounter{figure}{0}
\renewcommand{\thefigure}{S\arabic{figure}}
\renewcommand{\thepage}{S\arabic{page}}
\renewcommand{\thesection}{S\arabic{section}}
\renewcommand{\thetable}{S\arabic{table}}
\makeatletter

In this supplement material we show additional results and details to further solidify our findings in the main text. In particular, for the cubic discretization of momentum space we show the convergence of $\rho_{N_C}(0)$ and $\rho''_{N_C}(0)$ with the cut-off and the convergence of the peak of $\rho''_{N_C}(0)$ in the limit of large system size and KPM expansion order. We also show the quality of the fits for $\rho_{N_C}(0)$ (for the FCC discretization)  vs $N_C$ to extract $\rho(0)$ and $\rho''(0)$, and we further show saturation in system size for $\rho(0)$ and $\rho_{N_C}''(0)$ data presented in the main text. Lastly, we discuss the results of the self-consistent Born analysis to determine the cut-off dependence on the location of the perturbative transition, as this provides a qualitative framework to interpret cut-off dependence of the avoided quantum critical point.

Unless otherwise stated, all plots have error bars. In many cases, the data markers are larger than those error bars.

\section{The Kernel Polynomial Method}

The KPM uses a Chebyshev expansion such that for each realization, one can define $\tilde H = (H-b)/\Delta$ and $\tilde \epsilon = (\epsilon-b)/\Delta$ with $\Delta$ and $b$ chosen such that the spectrum of $\tilde H$ is within $(-1,1)$ and density of states is well approximated by
\begin{equation}
    \tilde \rho_{N_C}(\tilde \epsilon) = \frac1{\sqrt{1-\tilde \epsilon^2}}\left[g_0\mu_0 T_0(\tilde \epsilon) + 2 \sum_{n=1}^{N_C} g_n \mu_n T_n(\tilde \epsilon) \right]
\end{equation}
where $T_n$ are Chebyshev polynomials, $\mu_n = \tr[T_n(\tilde H)]$ (with the trace evaluated stochastically), and $g_n$ is the Jackson kernel which guarantees uniform convergence (suppression of Gibbs phenomena) as $N_C\rightarrow\infty$ and positivity of $\tilde \rho_{N_C}(\tilde \epsilon)$~\cite{Weisse-2006}.
This is then easily related back to $\rho_{N_C}(\epsilon) = \frac{1}{\Delta} \tilde \rho_{N_C}[(\epsilon - b)/\Delta ] $.

\section{Discretization data}

\subsection{Cubic data}

\subsubsection{Converging in cutoff}

While most of the rare region figures of merit seem to be cutoff-independent, the value of $\rho(0)$ as illustrated in the text is \emph{not}.
In order to demonstrate that this does not affect our results for larger cutoffs, we can test momentum space cutoff dependence of our data.
This is done explicitly with the cubic model.

\begin{figure}
    \centering
    \includegraphics[width=0.5\columnwidth]{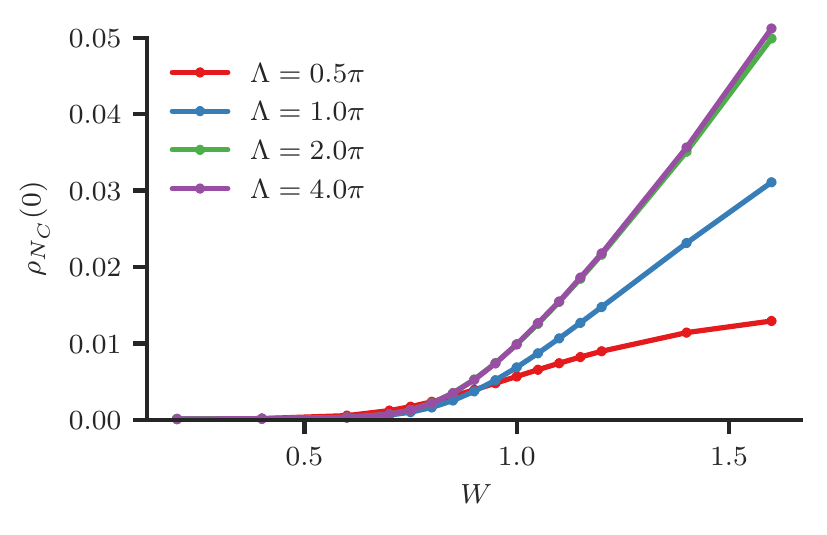}
    \includegraphics[]{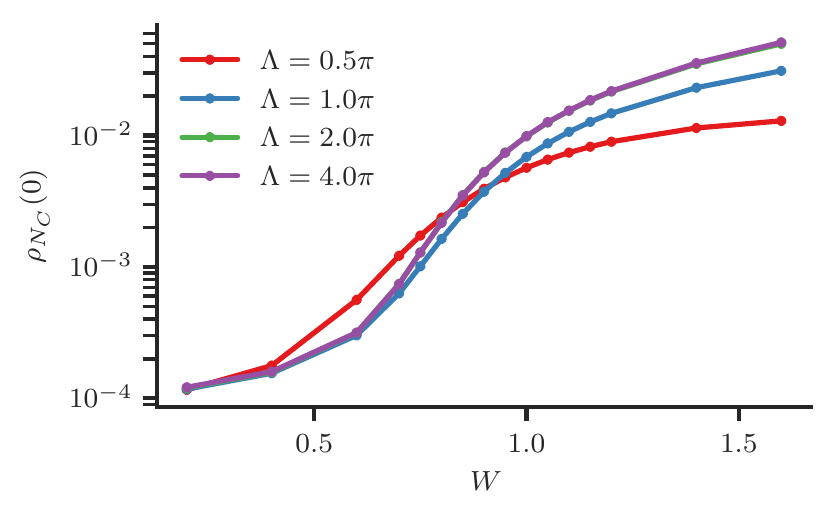}
    \caption{{\bf Cubic lattice.} (left) We can converge the density of states in cutoff. In this plot, we have used a system size $L = 32$ and an energy resolution $ \Lambda/N_C = \frac{\pi}{256}$. (right) Same data on a logarithmic scale.}
    \label{fig:cutoff_rho0data}
\end{figure}

First, we can determine that $\rho(0)$ does in fact converge in cutoff and a representative sample is shown for $L = 32$ in Fig.~\ref{fig:cutoff_rho0data}.


\begin{figure}
    \centering
    \includegraphics[width=0.5\columnwidth]{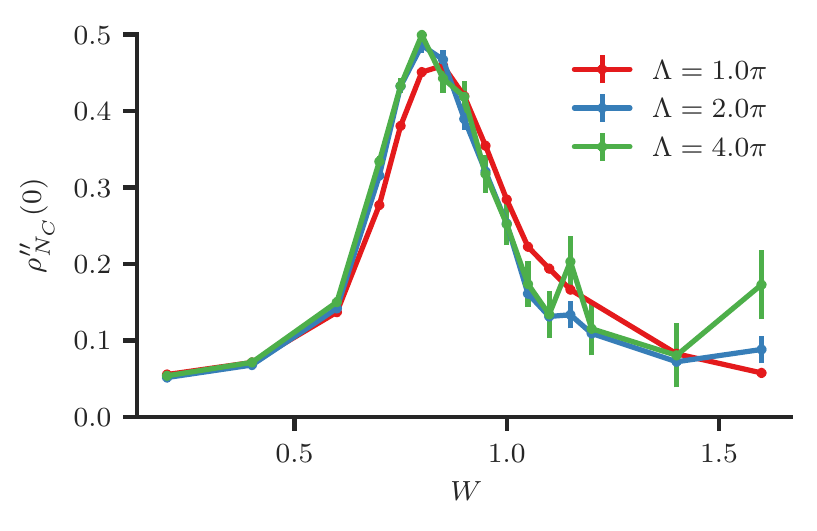}
    \caption{{\bf Cubic lattice.} Converged $\rho''(0)$ data for cutoff at a system size $L = 32$ and energy resolution $\Lambda/NC = \frac{\pi}{512}$.}
    \label{fig:cutoff_d2rho0data}
\end{figure}

Similarly, we can study $\rho''(0)$ vs.\ cutoff and we find that for higher energy resolutions, it is weakly affected as we can see in Fig.~\ref{fig:cutoff_d2rho0data}.
Importantly, it is appears to approach its infinite cutoff faster than $\rho(0)$ does.

\subsubsection{Converging in system size and energy resolution}

To converge the $\rho''(0)$ peak in both system size and energy resolution we pick a value of $N_C$ then increase the system size until it saturates.
If we increase $N_C$ and find it saturates to roughly the same value in $L$ we take  that to be converged.

\begin{figure}
    \centering
    \includegraphics{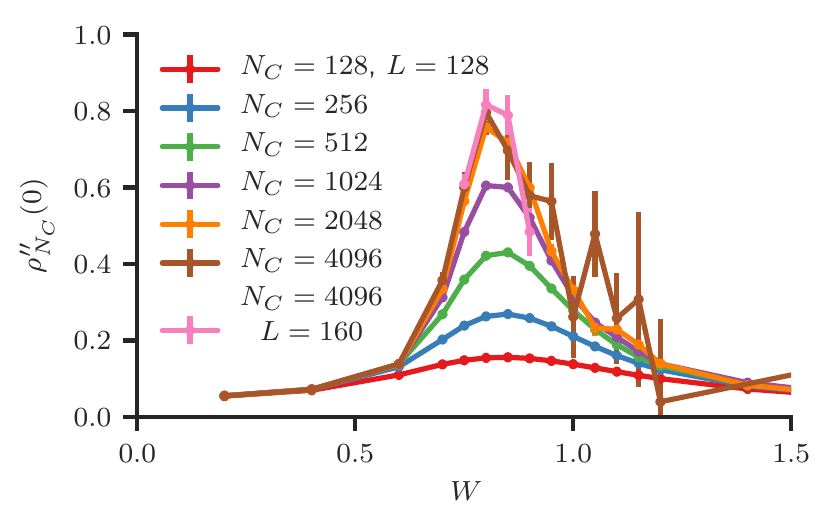}
    \caption{{\bf Cubic lattice.} The peak for cubic data at fixed cutoff $\Lambda = \pi$. The energy resolution is roughly given by $\Lambda /N_C$, and we have converged the peak. The resulting peak values are plotted in Fig 4(b) of the main text. }
    \label{fig:d2rho0_NCconverge}
\end{figure}

The result that we have for the cubic model is shown in Fig.~\ref{fig:d2rho0_NCconverge}.

\subsection{FCC data}

Most of the relevant FCC data appears in the main text. Here, we show the ``twist dispersion'' of the rare state, some of the fits that lead to Fig.~3 in the main text, and system size dependence of various quantities claimed in the main text to be saturated in system size.

\emph{Identifying the rare state:} The rare state is found by setting $\bm\varphi = (0.5,0.5,0.5)$ at $W=0.7$, and we can in fact verify the state appears localized by varying this quantity (which acts just as twisting the boundary conditions). At $\bm\varphi = (0.5,0.5,0.5)$, the system is time-reversal symmetric and we verify this numerically, leading to a Kramer's degeneracy.

\begin{figure}
    \centering
    \includegraphics[width=0.49\columnwidth]{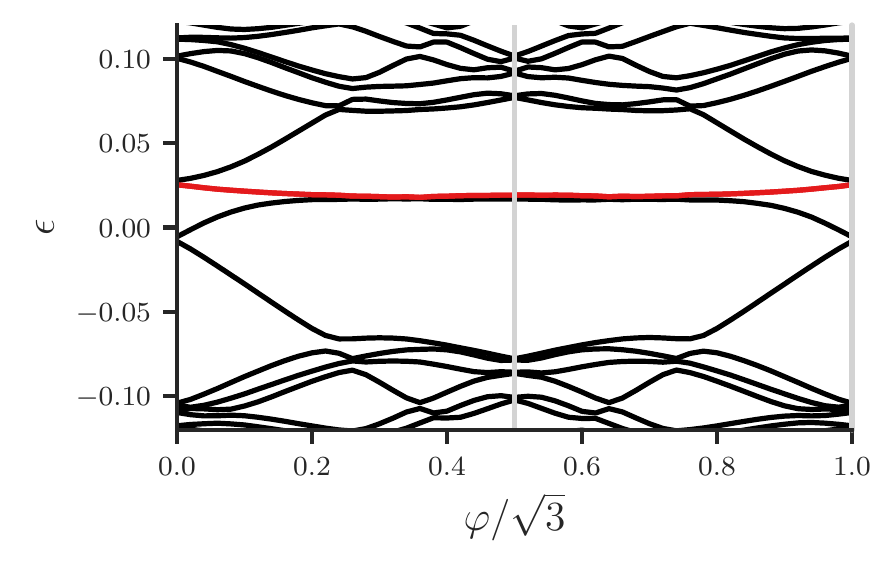}
    \includegraphics[width=0.49\columnwidth]{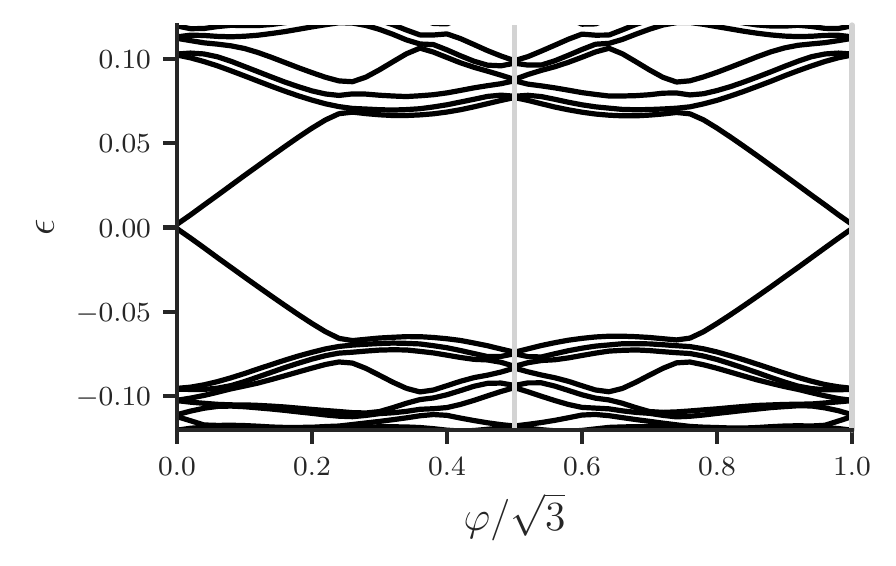}
    \caption{{\bf FCC lattice.} (left) The twist dispersion of the rare state pictured in Fig.~2 of the main text. The red represents the rare state. $\varphi = |\bm\varphi|$ which is varied such that $\bm \varphi = \varphi(1,1,1)/\sqrt{3}$. This state is found with $\Lambda = \pi/\sqrt{2}$, $L = 25\sqrt{3}$ and $W = 0.7$ (setting $\xi=1$ without loss of generality). (right) A sample with the same parameters, but no rare state.}
    \label{fig:twistdispersion}
\end{figure}

Varying $\bm\varphi$ amounts to  going through a cut in a mini-Brillioun zone, and we get what is pictured in Fig.~\ref{fig:twistdispersion}, which demonstrates a clear distinction between a rare weakly dispersing state and a perturbative Weyl state.
Note that there is a second flat, rare state. These two states are nearly identical in $|\psi(x,y,z)|$, differ slightly in energy and have different spinor structure.

\begin{figure}
    \centering
    \includegraphics[width=0.49\columnwidth]{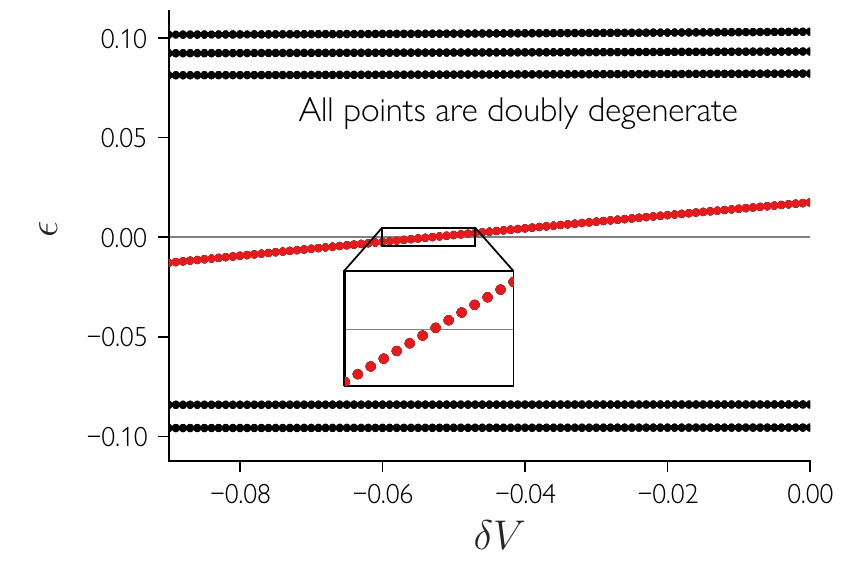}
    \caption{{\bf FCC lattice.} Tuning the (pair of) rare state(s) continuously through zero energy by slightly modifying the disorder potential with a small spherical potential centered around the maximal value of the rare state. We take the same potential as in Fig.~\ref{fig:twistdispersion}, but we truncate it down to size $L=24\sqrt{3}$ to guarantee time-reversal symmetry. The spherical potential has radius two, strength $\delta V$ and zero mean over all of real space. This is done at the time-reversal symmetric point $\bm \varphi = (0.5,0.5,0.5)/\sqrt{3}$ with the same rare state pictured in Fig.~\ref{fig:twistdispersion}(left) and Fig.~2 of the main text. As stated, all points are doubly degenerate.}
    \label{fig:tune_rare}
\end{figure}

While finding a rare state precisely at $E=0$ is nearly impossible to find with random number generation, we can roughly figure out how much we need to perturb our potential to move our rare state through $E=0$.
To accomplish this, we choose the realization that gives us Fig.~2 in the main text, we then add to an additional, small, spherical potential of radius 2
\begin{equation}
    \Delta V(\mathbf r) = \delta V \Theta(2 - |\mathbf{r} - \mathbf{r}_\mathrm{rare}|) - \braket{\delta V \Theta(2 - |\mathbf{r}' - \mathbf{r}_\mathrm{rare}|)}_{\mathbf r'},
\end{equation}
where $\mathbf r_{\mathrm{rare}}$ is where the magnitude of the rare state achieves its maximal value, and $\braket{\cdots}_{\mathbf{r'}}$ indicates an average over real space (this term ensures there is no constant shift; it vanishes in the limit of infinite size).
Once added, we see that we can perturb our rare state through $E=0$ with a relatively small $\delta V\approx -0.05$ (see Fig.~\ref{fig:tune_rare}). The nearby states remain relatively unaffected by this local perturbation.

\begin{figure}
    \centering
    \includegraphics{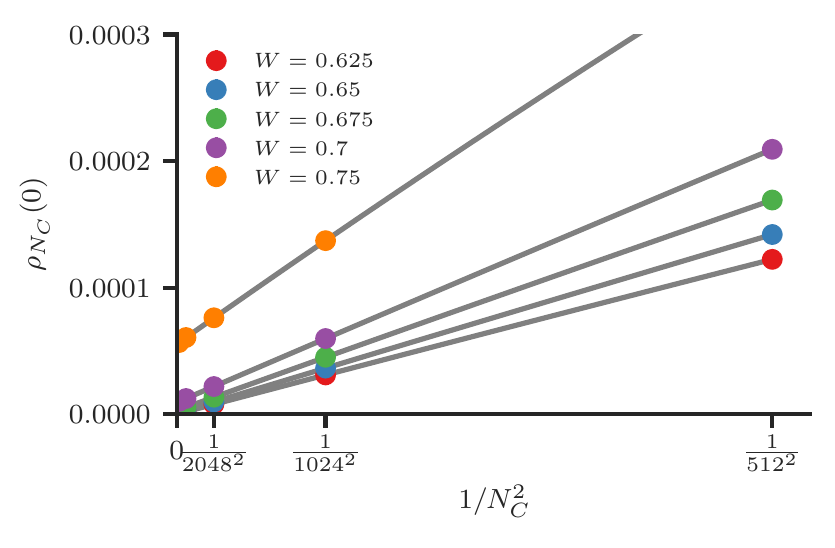}
    \includegraphics{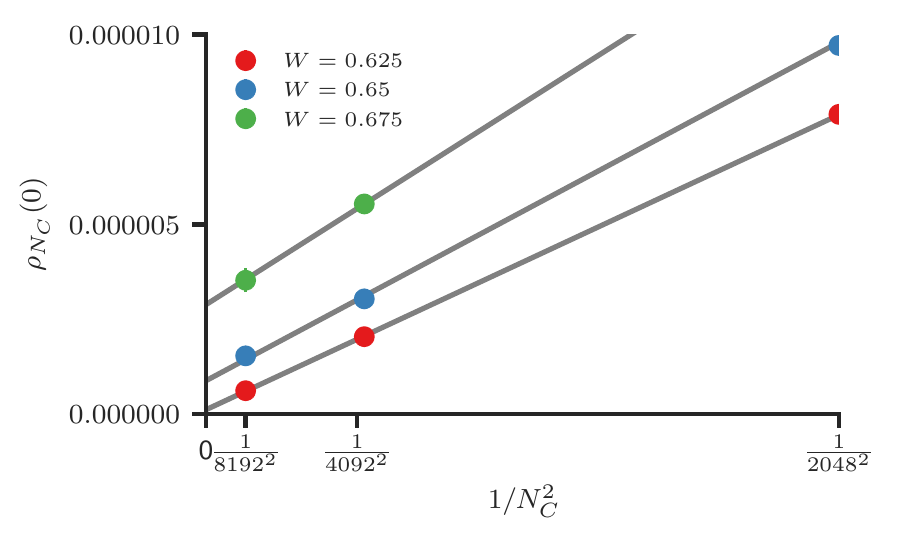}
    \caption{{\bf FCC lattice.} Fits of the $\rho(0)$ data with $N_C$. We fit the data according to Eq.~\eqref{eq:rho0NCfitting} to extract $\rho(0)$ and $\rho''(0)$. Both plots depict the same fits (gray lines), but on different scales. This data is for $L= 160\sqrt{3}$ and $\Lambda = \pi/\sqrt{2}$ on the FCC lattice.}
    \label{fig:rho0_FCC_fits}
\end{figure}

\emph{KPM and $N_C$ scaling:} For the fitting procedure we assume for the purposes of fitting
\begin{equation}
    \rho_{N_C}(0) = \rho(0) + \tfrac12 \rho''(0) \left( \frac{\pi \Delta }{N_C}\right)^2 +  \frac{c}{N_C^4},
    \label{eq:rho0NCfitting}
\end{equation}
where $\Delta$ is the bandwidth of the model and $c\propto \rho^{(4)}(0)$ (the fourth derivative with energy). Using a computed average bandwidth $\Delta$ we obtain a number of fits; a representative case is shown in Fig.~\ref{fig:rho0_FCC_fits} from before the avoided critical point. The fits are clearly consistent with a finite density of states in the limit $N_C \rightarrow \infty$.

\begin{figure}
    \centering
    \includegraphics{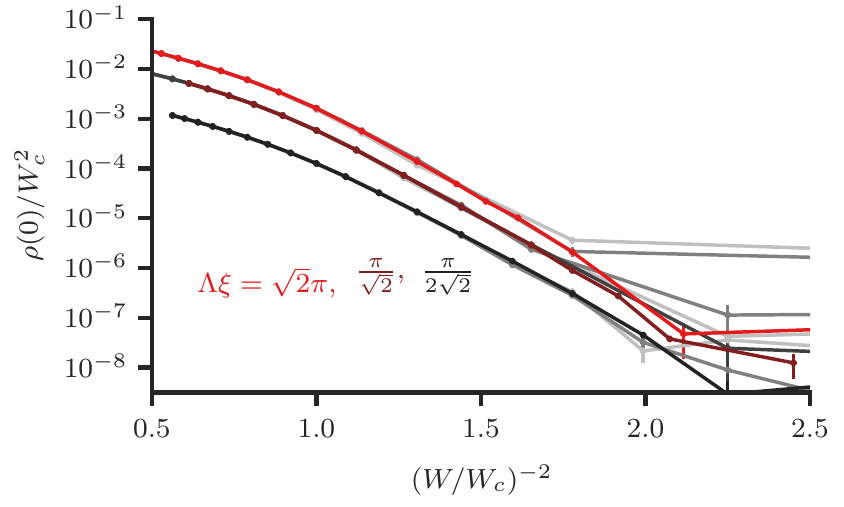}
    \caption{{\bf FCC lattice.} A plot of Fig.~3(c) in text but with smaller $L$ values plotted as well . All gray curves (light to dark) correspond to  $L=64\sqrt{3},128\sqrt{3},256\sqrt{3}$ for $\Lambda = \pi/2\sqrt{2}$, $L = 32\sqrt{3},64\sqrt{3},128\sqrt{3},160\sqrt{3}$ for $\Lambda = \pi/\sqrt{2}$  and $L=16\sqrt{3},32\sqrt{3},64\sqrt{3}$ for $\Lambda = \sqrt{2}\pi$.  For larger values of $\rho(0)$, all $L$ curves line up for each $\rho(0)$ making them difficult to distinguish. This shows a clear saturation of $\rho(0)$ in system size.}
    \label{fig:Lsaturatedrho0fit}
\end{figure}

In addition, we show the system size dependence of $\rho(0)$ in Fig.~\ref{fig:Lsaturatedrho0fit} demonstrating it is clearly converged in this range of $W$  below the AQCP.

\begin{figure}
    \centering
    \includegraphics{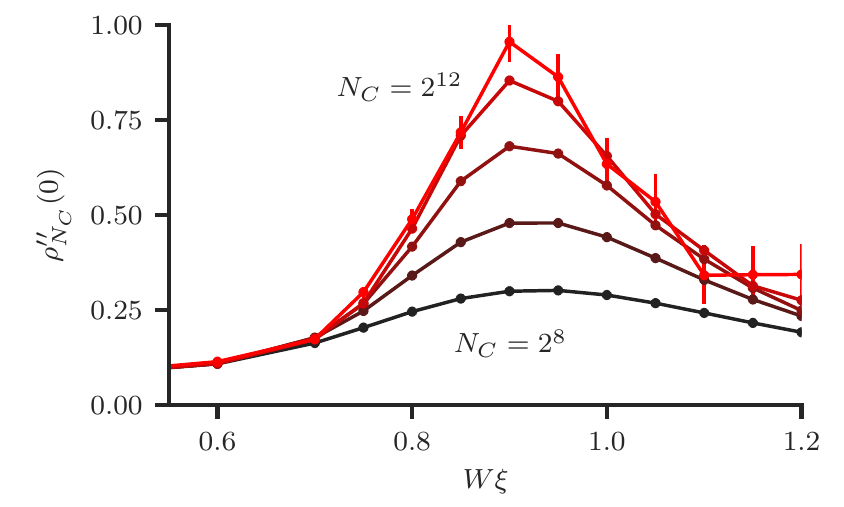}
    \includegraphics[]{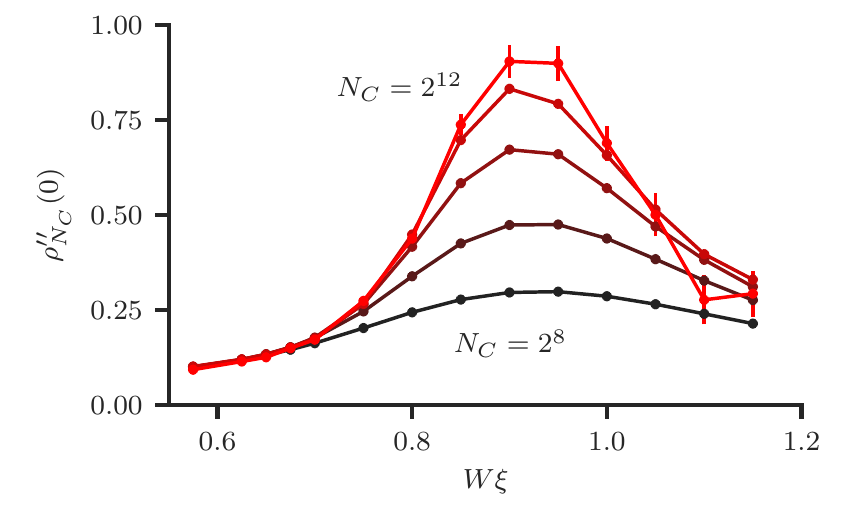}
    \caption{{\bf FCC lattice.} All data in this figure has $\Lambda = \pi/\sqrt{2}$. (left) $L=128\sqrt{3}$ data for $\rho_{N_C}''(0)$. (right) $L=160\sqrt{3}$ data for $\rho_{N_C}''(0)$. Notice that each $N_C$ is converged in system size.}
    \label{fig:d2rho0_NCdata_sizeconv}
\end{figure}

Finally, we also show how the data presented in Fig.~4(a) of the main text is converged in system size by comparing two system sizes $L=128\sqrt{3}$ and $L=160\sqrt{3}$ in  Fig.~\ref{fig:d2rho0_NCdata_sizeconv}.
Within error bars, they are nearly identical.

\section{Self-consistent Born approximation}

To get an idea of what the transition looks like for a finite cutoff, we turn to the self-consistent Born approximation.
In this situation, we look at the disorder averaged Green's function
\begin{equation}
    G = \frac{1}{i\omega - i\bm\sigma \cdot \bm \nabla - \Sigma} = \left\langle\frac{1}{i\omega + i\bm\sigma \cdot \bm\nabla - V(\mathbf r)}\right\rangle
\end{equation}
At second order, we can identify the self energy
\begin{equation}
    \Sigma = \left\langle V(\mathbf r) \frac{1}{i\omega - i\bm\sigma \cdot \bm \nabla - \Sigma} V(\mathbf r)\right\rangle.
\end{equation}
We can easily evaluate this in momentum space, where  $\Sigma$ is  diagonal (due to the disorder average), such that we have
\begin{align}
\Sigma & = \Sigma(i\omega, \mathbf k) (2\pi)^3\delta(\mathbf k - \mathbf k')\nonumber  \\ & = \int \frac{d^3 \mathbf k''}{(2\pi)^3} \frac{\braket{V(\mathbf k - \mathbf k'')V(\mathbf k''-\mathbf k')}}{i\omega - \mathbf k''\cdot \bm \sigma - \Sigma(i\omega, \mathbf k'')}.
\end{align}
The average can be evaluated from the correlator $\braket{V(\mathbf r)V(\mathbf R + \mathbf r)} = W^2 f(R/\xi)$ and
\begin{equation}
  \begin{split}
  \braket{V(\mathbf k) V(\mathbf k')} & = \int d^3 \mathbf r d^3\mathbf r'\braket{V(\mathbf r)V(\mathbf r')} e^{-i\mathbf k\cdot \mathbf r - i \mathbf k' \cdot \mathbf r'} \\
  & = \int d^3 \mathbf r d^3 \mathbf R\braket{V(\mathbf r)V(\mathbf r+\mathbf R)} e^{-i(\mathbf k' + \mathbf k)\cdot \mathbf r - i \mathbf k'\cdot \mathbf R} \\
  & = W^2 \int d^3 \mathbf r d^3 \mathbf R f(R/\xi) e^{-i(\mathbf k' + \mathbf k)\cdot \mathbf r - i \mathbf k'\cdot \mathbf R}\\
  & = W^2 \xi^3 f(\xi k)(2\pi)^3 \delta(\mathbf k+\mathbf k') .
\end{split}
\end{equation}
From which we can derive
\begin{equation}
  \Sigma(i\omega,\mathbf k) = W^2 \int \frac{d^3 \mathbf k'}{(2\pi)^3}  \frac{ \xi^3 f(\xi |\mathbf k - \mathbf k'|)}{i\omega - \mathbf k'\cdot \bm \sigma - \Sigma(i\omega,\mathbf k')}.
\end{equation}
Now, we set $\mathbf k=0$ so that we can further write this expression
\begin{equation}
  \Sigma(i\omega )= W^2 \int \frac{d^3 \mathbf k}{(2\pi)^3}  \frac{ \xi^3 f(\xi k)}{i\omega - \mathbf k\cdot \bm \sigma - \Sigma(i\omega)}.
\end{equation}
We can then simplify things a fair bit
\begin{equation}
  \Sigma(i\omega) = W^2 [i\omega - \Sigma(i\omega)] \int \frac{d^3\mathbf k}{(2\pi)^3} \frac{\xi^3 f(\xi k)}{(i\omega - \Sigma(i\omega))^2 - k^2}.
\end{equation}
If we concentrate on $\omega = 0$, then we have
\begin{equation}
  \Sigma(0) = W^2 \Sigma(0) \int \frac{d^3\mathbf k}{(2\pi)^3} \frac{\xi^3 f(\xi k)}{-\Sigma(0)^2 + k^2},
\end{equation}
and with spherical symmetry
\begin{equation}
  \Sigma(0) = \frac{W^2 \xi^3}{2\pi^2} \Sigma(0) \int_0^\infty dk \frac{ k^2  f(\xi k)}{-\Sigma(0)^2 + k^2}.
\end{equation}
This clearly has a solution of $\Sigma(0) = 0$ which is the uninteresting semimetal, we are interested in when the other root passes zero
\begin{equation}
  1 = \frac{W_c^2 \xi^3}{2\pi^2} \int_0^\infty dk  f(\xi k).
\end{equation}
Now, let us consider $f(x) =  e^{-x^2}$, then $f(k) = \pi^{3/2} e^{-k^2 /4}$, and $\int_0^\infty  f(\xi k)  dk =  \pi^2/\xi  $
\begin{equation}
  W_c\xi = \sqrt{2}, \quad \text{Infinite cutoff}.
\end{equation}
We can introduce a cutoff at this point to find where the transition is. It would correspond to putting $\int_0^\Lambda dk \tilde f(\xi k) =  \pi^2 \erf(\Lambda \xi/2)/\xi$, so that we have
\begin{equation}
  W_c\xi = \sqrt{\frac{2}{\erf(\Lambda\xi/2)}}, \quad \text{Finite cutoff.} \label{eq:WcContinuumCutoff}
\end{equation}
Qualitatively, this captures the critical point at smaller cutoffs, as the blue curve indicates in Fig.~\ref{fig:scba}.

\subsection{Lattice regularization}

\begin{figure}
  \includegraphics[width=0.5\columnwidth]{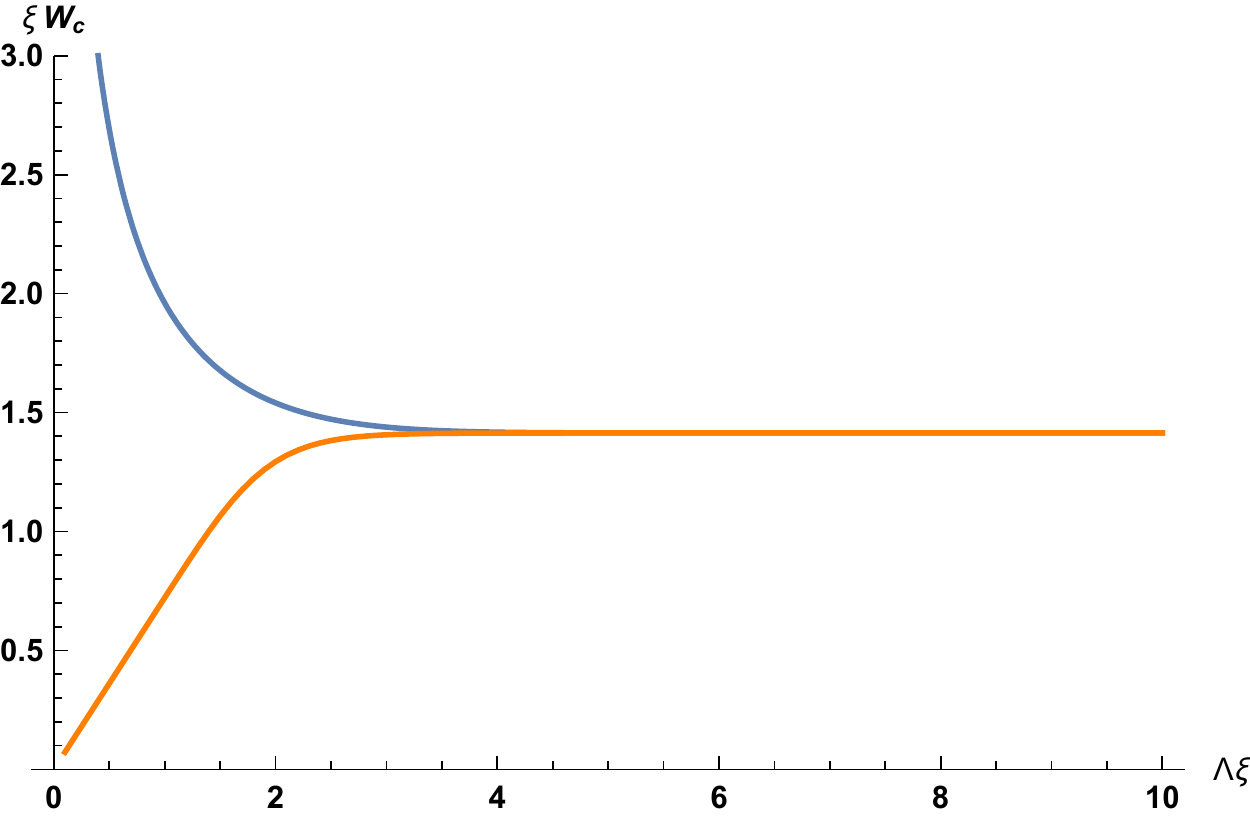}
  \caption{The critical $W_c$ vs cutoff as computed by the self-consistent Born approximation. The blue curve diverges at small $\Lambda\xi$ and represents use of the continuum computed $f(\xi k)$ while the orange curve computes the fourier transform of $f(R/\xi)$ on a lattice. The lattice regularized version vanishes at small $\Lambda \xi$ consistent with a white noise theory where disorder correlation length is controlled entirely by the cutoff. This is a plot of analytic functions and thus has no error bars.}
  \label{fig:scba}
\end{figure}

In the case of the cubic lattice, we use the lattice itself to implement the cutoff.
What we mean by this is that instead of using $f(k\xi)$ as computed in the continuum, we compute the Fourier transform explicitly
\begin{equation}
\begin{split}
    f(ka,\xi/a) & = a^3 \sum_{\mathbf n} e^{-n^2 a^2/\xi^2} e^{i\mathbf k \cdot \mathbf n a} \\
     & = a^3 \prod_{i=x,y,z} \vartheta_3(k_i a/2, e^{-a^2/\xi^2}),
\end{split}
\end{equation}
where $\vartheta_3$ is the Jacobi elliptic function.
Doing the same analysis as the previous section, we get
\begin{equation}
    W_c\xi = \frac{\Lambda \xi}{\pi} \sqrt{\frac1{g(\Lambda \xi)}},
\end{equation}
for $\Lambda = \pi/a$ and with
\begin{equation}
    g(\Lambda \xi) = \int_{-\pi}^\pi \frac{d^3 y}{(2\pi)^3}  \frac{\prod_{i=x,y,z} \vartheta_3(y_i/2, e^{-\pi^2/(\Lambda\xi)^2})}{y^2}.
\end{equation}
This leads to different small cutoff behavior for the critical $W_c$ as shown by the line that vanishes at $\Lambda \xi = 0$ in Fig.~\ref{fig:scba}.

\end{document}